\documentclass[useAMS,usenatbib,usegraphicx]{mn2e}

\usepackage{hyperref}
\usepackage{color}

\title[CI Aql: a Type Ia Supernova progenitor?]{CI Aql: a Type Ia supernova progenitor?}
\author[D. I. Sahman et al]{D. I. Sahman,$^{1}$\thanks{E-mail: 
d.sahman@sheffield.ac.uk (DIS); vik.dhillon@sheffield.ac.uk (VSD)} V. S. Dhillon,$^{1}$\footnotemark[1]
T. R. Marsh,$^{2}$
S. Moll,$^{1}$
T. D. Thoroughgood,$^{1}$
\newauthor
C. A. Watson$^{3}$
and S. P. Littlefair$^{1}$\\
$^{1}$Department of Physics and Astronomy, University of Sheffield, Sheffield S3 7RH, UK\\
$^{2}$Department of Physics, University of Warwick, Coventry CV4 7AL, UK\\
$^{3}$Astrophysics Research Centre, School of Mathematics \& Physics, Queen's University Belfast, University Road, Belfast BT7 1NN, UK\\}

\begin{document}

\date{Accepted 2013 May 9. Received 2013 April 30; in original form 2013 March 5}

\pagerange{\pageref{firstpage}--\pageref{lastpage}} \pubyear{2013}

\maketitle

\label{firstpage}

\begin{abstract}

If recurrent novae are progenitors of Type Ia supernovae, their white dwarfs must have masses close to the Chandrasekhar limit. The most reliable means of determining white dwarf masses in recurrent novae is dynamically, via radial-velocity and rotational-broadening measurements of the companion star. Such measurements require the system to be both eclipsing and to show absorption features from the secondary star. Prior to the work reported here, the only dynamical mass estimate of a recurrent nova was for U Sco, which has a white dwarf mass of $1.55 \pm 0.24$ M$_{\sun}$
\citep{thoroughgood01}.

We present new time-resolved, intermediate-resolution spectroscopy of the eclipsing recurrent nova CI Aquilae (CI Aql) during quiescence. We find the mass of the white dwarf to be $1.00 \pm 0.14$ M$_{\sun}$ and the mass of the secondary star to be $2.32 \pm 0.19$ M$_{\sun}$.  We estimate the radius of the secondary to be $2.07 \pm 0.06$ R$_{\sun}$, implying that it is a slightly-evolved early A-type star. The high mass ratio of $q=2.35 \pm 0.24$ and the high secondary-star mass implies that the mass transfer occurs on a thermal timescale. We suggest that  CI Aql is rapidly evolving into a supersoft X-ray source, and ultimately may explode as a Type Ia supernova within 10 Myr.
\end{abstract}

\begin{keywords}
binaries: eclipsing -- binaries: spectroscopic -- stars:
individual stars: CI Aql -- novae, cataclysmic variables -- supernovae: general.
\end{keywords}

\section{Introduction}

Type Ia supernovae (SNIa) play a key role in modern cosmology. The rate of decline of their light curves and their intrinsic luminosities are directly correlated 
\citep{phillips93}
and so they provide an ideal standard candle to measure cosmological distances more reliably than other techniques currently available. The discovery of the accelerating expansion of the Universe by
\citet{riess98}
 and
\citet{perlmutter99}
was made using SNIa. However, despite over thirty years of research, their progenitors are not yet satisfactorily explained. The canonical view is that SNIa are due to the thermonuclear explosion of a near--Chandrasekhar mass ($M_\mathrm{Ch}$) carbon-oxygen (CO)  white dwarf, which has accreted mass from a companion star. It is the nature of the white dwarf's companion that remains elusive.

 The two favoured models are the double degenerate model (DDM) where the companion is another white dwarf
  (\citealt{webbink84};
\citealt{iben84}),
 or the single degenerate model (SDM) where the companion is non-degenerate  e.g. a main sequence star, a helium star or a red giant 
 \citep{whelan73}.
 In the DDM, two CO white dwarfs, with a combined mass greater than $M_\mathrm{Ch}$, lose angular momentum through the emission of gravitational waves. This causes their orbital separation to shrink, leading to merger and ultimately an explosion as a SNIa. In the SDM, the white dwarf accretes matter from its companion, either by Roche lobe overflow or by wind-accretion, and gradually approaches $M_\mathrm{Ch}$. Just prior to the point at which the white dwarf would undergo gravitational collapse, carbon fusion begins in the core of the white dwarf, and leads to a runaway thermonuclear explosion which disrupts the entire white dwarf. This would explain the homogeneity seen in the luminosities of SNIa. However, each model has its strengths and weaknesses.

The DDM can readily explain the lack of hydrogen seen in most SNIa spectra, as all the hydrogen has been processed by the time of the merger
(\citealt{hamuy03};
\citealt{livio03}). 
The DDM is also favoured by the non-detection of any confirmed progenitors or survivors of nearby SNIa
(\citealt{kerzendorf09};
\citealt{schaefer12}),
 and the lack of spectral features from the impact of the supernova forward shock with the secondary star
 \citep{badenes07}
 or the circumstellar material blown off the secondary prior to the explosion
 \citep{leonard07}.
 For example, the nearest SNIa in decades, SN2011fe in M101, has been observed at multiple wavelengths and at multiple epochs, and shows no trace of interactions with either circumstellar material or a progenitor, nor any evidence of a companion
 (\citealt{nugent11};
 \citealt{bloom12};
 \citealt{brown12};
 \citealt{horesh12}).
 This rules out most SDM channels for this event. However, there is evidence against the DDM from large--scale surveys looking at populations of white dwarfs (e.g. see
 \citealt {napiwotzki07};
\citealt{badenes12})
 which have detected insufficient numbers of high mass white dwarf pairs to account for the observed SNIa rate. Simulations of double white dwarf mergers also suggest that most mergers would lead to accretion induced collapse to form a neutron star
rather than SNIa explosions
(\citealt{shen12}).

Studies of the delay time distribution of SNIa since star formation show a bimodal distribution, suggesting that perhaps both the SDM and DDM channels contribute to the overall SNIa rate
\citep{maoz11}.
 Some SNIa seem to be derived from super--$M_\mathrm{Ch}$ white dwarfs, which favours the DDM, whereas other low-luminosity SNIa suggest they were derived from  sub-$M_\mathrm{Ch}$ white dwarfs
\citep{vankerkwijk10},
which neither model can easily account for, but which are better explained as a sub-class of the DDMs  comprising sub-$M_\mathrm{Ch}$ CO and He white dwarf pairs
(\citealt{sim12}). 
In these 'double detonation' models, the He-rich material is transferred to the CO white dwarf. As the temperature and pressure of the accreted layer build up, it detonates and the shock creates a detonation in the core of the CO white dwarf. 
\citet{pakmor13}
and
\citet{shen13} 
have found that these models can also explain other features of SNIa previously only explained by SDM channels.

A small number of SNIa spectra show variable sodium absorption lines, and the variability has been attributed to the impact with circumstellar material
\citep{simon09},
 thereby supporting the SDM, but
 \citet{shen12}
 give other DDM--based interpretations of these results. The SDMs require a certain amount of fine-tuning in order that the white dwarf can accumulate mass despite the mass-loss in nova events
(\citealt{hachisu99a}).  
However, support for this fine tuning has come from
\citet{schaefer11},
who measured a small change in the orbital period  of CI Aql across its 2000 eruption thereby showing a net increase in the mass of the white dwarf.

One of the more favoured candidates in the SDM are recurrent novae (RNe). RNe are a class of cataclysmic variable stars (CVs) that  have more than one recorded nova outburst -- see
\citet{anupama08} and
\citet{schaefer10}
for reviews. The recurrence of outbursts is attributed to the near--$M_\mathrm{Ch}$ mass of the white dwarf in these systems, which is why they are favoured SNIa progenitor candidates. This has been confirmed for the RN U Sco, which was found to have a white dwarf mass of $1.55 \pm 0.24$ M$_{\sun}$
\citep{thoroughgood01}.
This made U Sco the only dynamical (and hence reliable) white dwarf mass determination in an RN. The technique of dynamical mass determination requires the system to be both eclipsing and to show spectral features of the secondary star.\footnote{It is also possible to use ellipsoidal variations in non-eclipsing systems to determine the inclination, but this is less reliable due to the presence of flickering and other phase-dependent light variations, e.g. due to a bright spot.}
The discovery of the 2000 outburst of CI Aql drew attention to this RN system, as it too is eclipsing and the secondary is visible, making it an ideal candidate for further mass determination.

CI Aql was first recorded in outburst in 1917
 \citep{reinmuth}.
 \citet{duerbeck87a} 
 suggested that it was either a dwarf nova with a long cycle length or a nova whose maximum was missed. 
 The second recorded outburst of CI Aql was reported in 2000 May 
(\citealt{takamizawa00};
\citealt{yamaoka00}),
with a magnitude of $m_V\sim7.5$, which secured its place amongst the 10 known RNe. 
\citet{schaefer01}
 found archival evidence of another outburst in 1941. This led him to propose that CI Aql has an outburst recurrence time-scale of approximately 20 yr, with the 1960 and 1980 outbursts failing to have been observed.  However, the time-scales of RN outbursts often vary, and the low accretion rate prior to the 2000 outburst, implied from their light--curve analysis model, led
  \citet{lederle03}
   to suggest that there were no outbursts between those of 1941 and 2000. CI Aql shows eclipses on an orbital period of 14.8 h, and has a quiescent magnitude of $m_V\sim15.7$
\citep{lederle03}.  
The pre-outburst quiescent optical spectrum showed emission lines of helium and
the Bowen complex on a reddened continuum, with all Balmer lines in absorption
\citep{greiner96}.
\citet{lynch04}
took infrared spectra at eight epochs between days 3 and 391 from peak brightness of the 2000 outburst. They found that the spectra were very similar to other RNe, except for an overabundance of nitrogen lines, suggesting that there was mixing of the white dwarf material with the accreted matter. This implies that the ejecta contained dredged-up white dwarf material, weakening the hypothesis that only accreted matter is ejected during the nova event. 

The determination of accurate masses of the components of RNe is essential to constrain models of their evolution and to establish whether they are realistic progenitors of SNIa. In this paper, we present time-resolved spectroscopy of CI Aql to determine the radial velocity, rotational velocity of the companion star and orbital inclination of the system, and hence measure the masses of the two component stars.

\section{Observations and Data Reduction}

The data were recorded over three nights, from 2003 June 10--12, when CI Aql was in quiescence, with the 4.2m William Herschel Telescope (WHT) on La Palma. A full orbit was observed during this time, including a primary eclipse. The double-armed spectrograph ISIS (Intermediate dispersion Spectrograph and Imaging System;
\citealt{carter93})
was used with a 1 arcsec slit width. We obtained 184 spectra of CI Aql simultaneously in red and blue wavebands. For the red arm, the Marconi2 CCD (2148$\times$4700 pixels) was used, with the R1200R grating, giving a wavelength range of $\lambda\lambda6041$--$6877$\AA\,and a resolution of 43 km\,s$^{-1}$. The blue arm  was configured with the EEV12 CCD (2148$\times$4200 pixels)
and the R1200B grating, giving a wavelength range of $\lambda\lambda4570$--$5370$\AA\, and a resolution of 55 km\,s$^{-1}$. All exposures were 600s. A comparison star was placed on the slit to correct for slit losses. 
Spectra of several templates of luminosity classes III--V ranging from F8 to M0 spectral type were also taken. A flux standard star and telluric star were observed on each night to calibrate between observed counts and flux, and to remove atmospheric absorption lines from the spectra respectively. Arc frames were taken between every five CI~Aql spectra to calibrate the wavelength scale, and to account for spectrograph flexure at different telescope positions;  the root-mean-square error in the fourth-order polynomial fits to the arc lines was $\sim 0.009$\AA. Flat-field frames and bias frames were also taken to correct for detector artefacts during the data reduction. The seeing varied between 1 and 3 arcsec and the sky was photometric on all three nights except for cirrus for two hours on the second night and some cirrus early on the third night. The spectra were reduced using the standard procedures outlined by
 \citet{thoroughgood01}. 
 A journal of the observations is given in Table \ref{journal}.

\begin{table}
 \caption{Journal of observations for CI Aql. Orbital phase is calculated using the ephemeris presented in Section \ref{ephem}.}
\label{journal}
\begin{tabular}{@{}lccccc}
  \hline
 { \sevensize UT} date & {\sevensize UT} & {\sevensize UT} & No. of & Phase & Phase \\
  (at start of night) & start & end & spectra & start & end \\
   \hline
   2003 June 10 & $23\colon44$ & $05\colon02$ & $29$ & $0.37$ & $ 0.71$ \\
   2003 June 11 & $23\colon21$ & $05\colon09$ & $31$ & $1.95$ & $ 2.34$ \\
   2003 June 12 & $23\colon41$ & $05\colon37$ & $32$ & $3.58$ & $ 3.99$ \\  
      \hline
 \end{tabular}  
 \end{table}

\section{Results}

\subsection{Light curves}
\label{lcur}

\begin{figure}
  \vspace*{5pt}
\includegraphics[width=85mm]{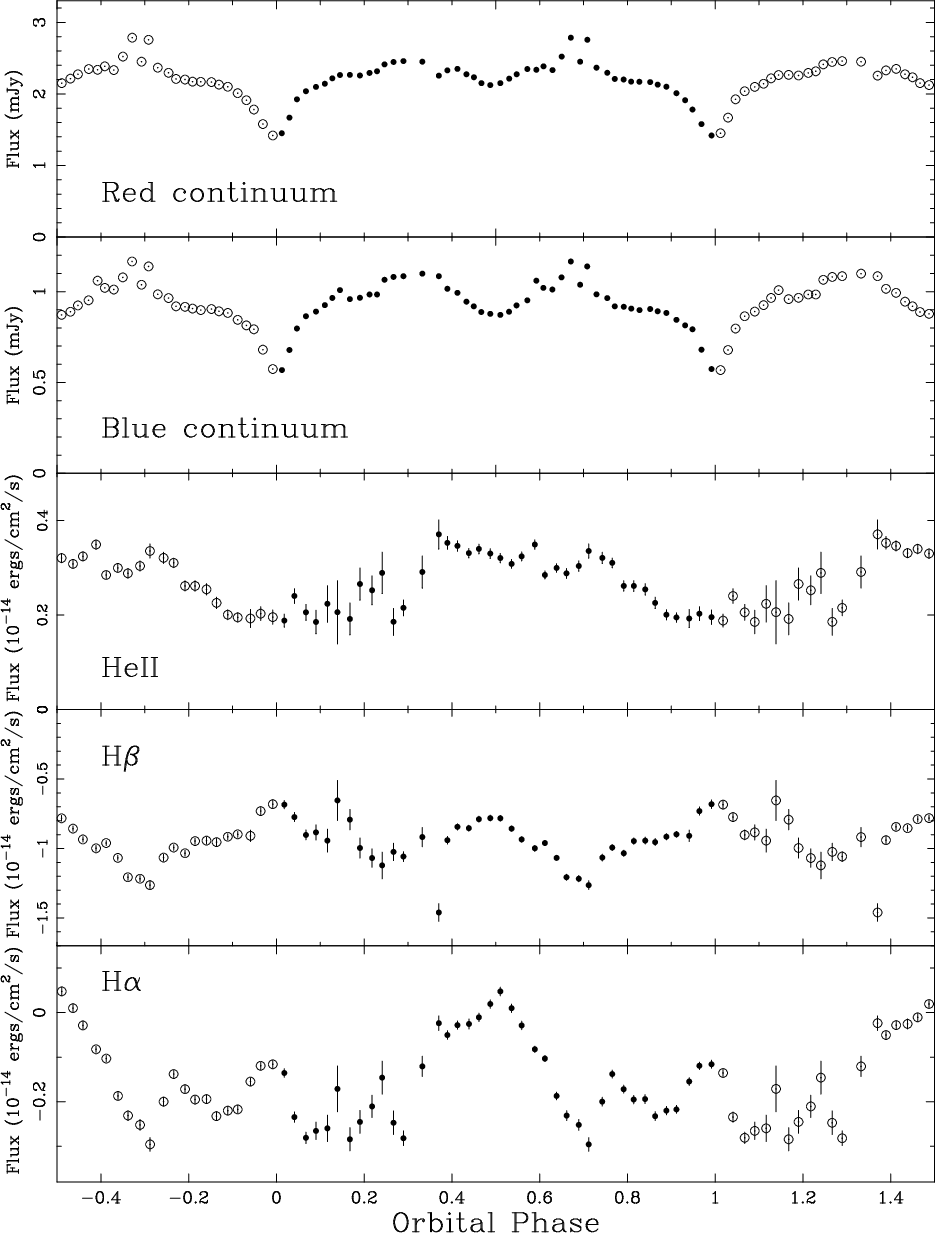}
  \caption{Continuum and emission-line light-curves of CI Aql. The data with open circles have been folded over from the real data (solid points) for clarity.}
  \label{lc}
\end{figure}

The slit-loss corrected light curves of CI Aql are presented in Fig. \ref{lc}. The red $(\lambda\lambda6079$--$6838$\AA) and blue ($\lambda\lambda4585$--$5319$\AA) continuum light curves show a primary eclipse of approximate depth 0.6 magnitude, and a dip of depth 0.1 magnitude at phase 0.5. The continuum light curves have two components, an eclipse of the white dwarf, and ellipsoidal modulation of the secondary, the latter being confirmed using the equations given by  
\citet{morris93}
and the coefficients for limb and gravity darkening of 
\citet{claret11},
in conjunction with the parameters derived in Section \ref{par}.

\citet{schaefer10}
 notes that there is flickering present throughout the light curve, except during the eclipse. This may suggest that the flickering originates in the disc. Our light curve shows flickering around phase 0.7, which is consistent with this picture.
 
 The middle panel of Fig. \ref{lc} shows the emission light curve for He {\sevensize II}, which exhibits a broad eclipse-like feature that appears to begin earlier and end later than the continuum eclipse. This is consistent with an origin on the inner hemisphere of the secondary star.
 
The lower two panels show the light curves for the H$\beta$ and H$\alpha$ lines. The H$\beta$ and H$\alpha$ lines are in absorption, although H$\alpha$ goes into emission around phase 0.5. Both light curves show weakest absorption at phases 0 and 0.5, suggesting that the absorption originates on the inner hemisphere of the secondary whilst the emission originates from the white dwarf or the disc.

 \subsection{Ephemeris and orbital period}
 \label{ephem}

The time of mid-eclipse was determined by fitting a Gaussian to the eclipse minimum of the continuum light curves in Fig. \ref{lc}, giving a value of $T_{mid-eclipse}$ = HJD  2452802.5145(3). This zero-point was used in conjunction with the orbital period reported in
\citet{lederle03},
$P$ =  0.6183634(3) 
{\color{blue}day},
to place our data on a phase scale. Note that 
\citet{schaefer11}
used data from 80 eclipse timings to derive a period of $P$ = 0.6183609(5) 
{\color{blue}day},
which is not sufficiently different to affect the results presented here.
\citet{schaefer11}
also found no evidence for  a non-zero value of $\dot{P}$.
 
 \subsection{Average spectrum}
 \label{avg}

\begin{figure}
 \vspace{10pt}
\includegraphics[width=66mm,angle=270]{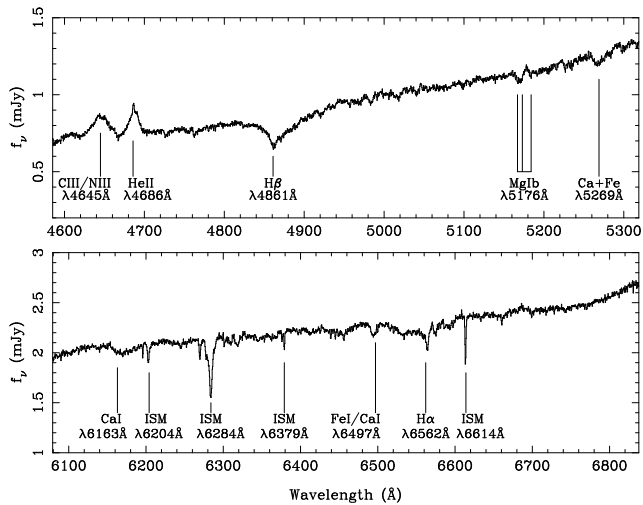}
 \caption{The average blue (top)  and red (bottom) spectra 
 {\color{blue}of}
  CI Aql, uncorrected for orbital motion.}
 \label{av}
\end{figure}

The average blue and red spectra are shown in Fig. \ref{av}. Note that the spectra have not been corrected for orbital motion, and so the weaker spectral features will be smeared out. Absorption features from the secondary star, such as Mg {\sevensize I}b, Ca~{\sevensize I} and Fe {\sevensize I} are visible, as well as high--excitation emission features such as He {\sevensize II} and the C {\sevensize III}/N {\sevensize III} complex. The Balmer lines are in absorption but our later analysis (see Section \ref{trail}) will show  them to be a combination of absorption and emission components. In Table \ref{flux} we list the fluxes, equivalent widths and velocity widths of the more prominent lines in the average spectra.

\begin{table*}
 \centering
  \begin{minipage}{110mm}
 \caption[c]{Fluxes and widths of prominent lines in CI Aql,  measured from the average spectrum.}
 \label{flux}
  \begin{tabular}[c]{@{}lrrrl@{}}
  \hline
   Line     & Flux $\times 10^{-14}$\,\, & EW\,\,\,\,\,\,\,\,\,\, & FWHM\,\,\,\, &    \,\,\,\,\,\,\,FWZI \\
    & (erg cm$^{-2}$ s$^{-1}$) & (\AA)\,\,\,\,\,\,\,\,\,\, & (km\,s$^{-1}$)\,\, & \,\,\,\,\,(km\,s$^{-1}$) \\
 \hline
 He  {\sevensize II} $\lambda4686$\AA$$ & 3.07 $\pm $ 0.04 & 3.07 $\pm $ 0.04 &  860 $\pm$ 100 & 1720 $\pm $ 200 \\
 C {\sevensize III}/N {\sevensize III} $\lambda4645$\AA$$ & 3.91 $\pm $ 0.05 & 3.80 $\pm $ 0.05 & 1500 $\pm $ 300 & 2790 $\pm $ 400 \\
 H$\beta$ & $-$11.05 $\pm$ 0.08 & $-$10.02 $\pm$ 0.07 & 1700 $\pm$ 400 & 6700 $\pm$ 1000 \\
\hline
\end{tabular}
\end{minipage}
\end{table*}

\subsection{Trailed spectra and Doppler tomography}
\label{trail}

\begin{figure}
  \vspace*{10pt}
  \centering
\includegraphics[width=63mm,angle=270]{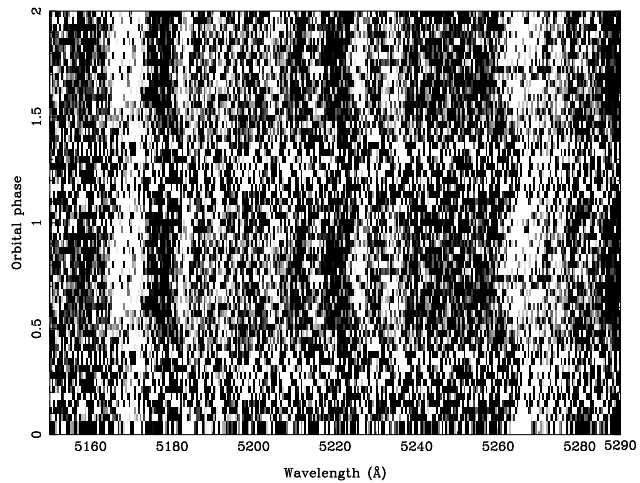}
  \caption{Trailed 
  {\color{blue}spectra}
   of the $\lambda\lambda5150-5290$\AA\ region, showing absorption features {\color{blue} (lighter grey-scales)} of Mg {\sevensize I}b, Ca {\sevensize I} and Fe {\sevensize I} from the secondary. The data have been folded in orbital phase for clarity. Note that the noisier data visible around phases 0.1--0.4 are due to a period of cloud. This is also evident in Figs 4, 6, 9 \& 10.}
  \label{tr1}
\end{figure}

The trailed spectrum of the  $\lambda\lambda5150$--$5290$\AA\ region is shown in Fig. \ref{tr1}, highlighting absorption features of Mg {\sevensize I}, Ca {\sevensize I} and Fe {\sevensize I}. Note that the increased noise around phases 0.1--
{\color{blue}0.4}
 is due to clouds. The absorption lines show a clear sinusoidal motion, which is red-shifted after phase 0 and crosses to blue-shifted at phase 0.5, indicating that they originate on the secondary. This part of the spectrum is used in the determination of the radial velocity of the secondary (see Section \ref{rad})

\begin{figure}
  \vspace*{10pt}
  \centering
\includegraphics[width=70mm,angle=270]{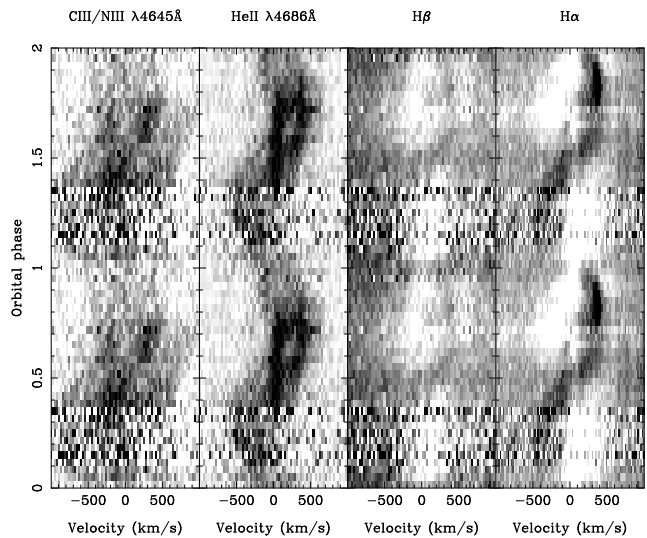}
  \caption{Trailed spectra of the C {\sevensize III}/N {\sevensize III} $\lambda4645$\AA, He {\sevensize II} $\lambda4686$\AA, H$\beta$  and H$\alpha$ lines,  with emission/absorption shown in dark/light grey-scales. The data have been folded in orbital phase for clarity.}
  \label{tr}
\end{figure}

The trailed spectra of the C {\sevensize III}/N {\sevensize III}, He {\sevensize II}, H$\beta$ and H$\alpha$ lines are shown in Fig. \ref{tr}. The C {\sevensize III}/N {\sevensize III} and He {\sevensize II} trails show sinusoidal motion, which is blue-shifted after phase 0 and crosses to red-shifted at phase 0.5.  This indicates that the lines originate from the disc or the white dwarf. The He {\sevensize II} trail shows two components, a narrow approximately constant velocity feature centred on zero velocity and a sinusoidal feature. The constant-velocity feature must be located very close to the centre of mass of the system, we explore the source of this feature below. The H$\beta$ trail is dominated by an absorption feature that moves in phase with the Mg {\sevensize I}b, Ca {\sevensize I} and Fe {\sevensize I} features in Fig. \ref{tr1}, and so most likely originates on the secondary. 
The H$\alpha$ trail shows two distinct elements moving in anti-phase to each other. There is a strong emission feature (dark grey-scales) associated with the disc and an absorption feature (light grey-scales) from the secondary.

\begin{figure}
  \vspace*{10pt}
\includegraphics[width=150mm,angle=270]{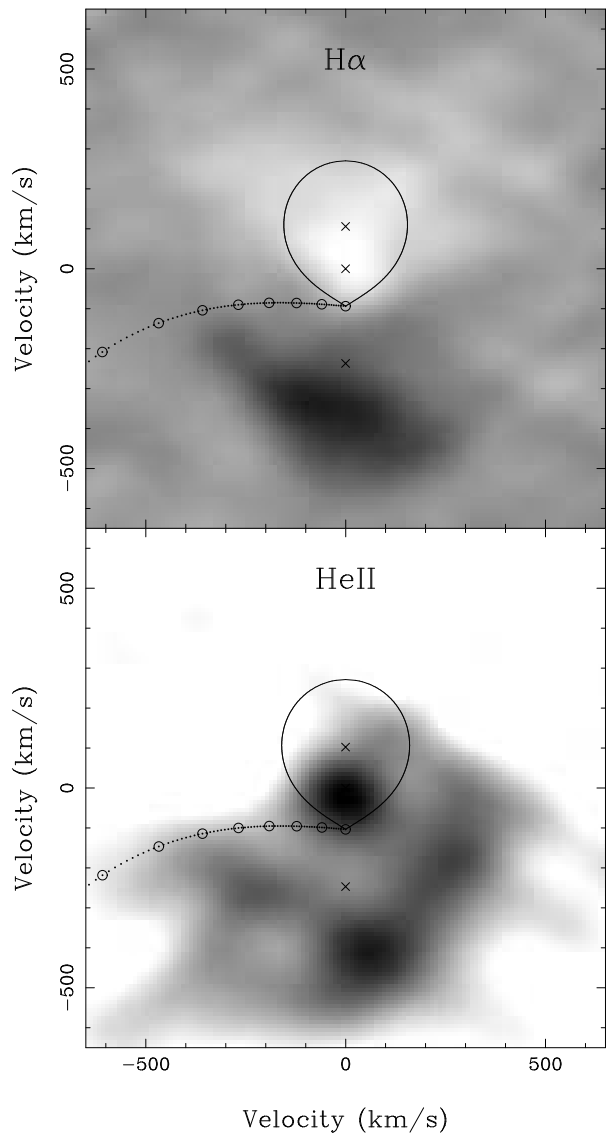}
  \caption{Doppler map of H$\alpha$ (top) and He {\sevensize II} $\lambda4686$\AA\ (bottom). The three crosses  represent the centre of mass of the secondary (upper cross), the centre of mass of the system (middle cross), and the centre of mass of the white dwarf (lower cross). The Roche lobe of the secondary star and the predicted trajectory of the gas stream have been plotted using the mass ratio, $q$=$M_2/M_1$ = 2.35 derived in Section \ref{par}. The series of circles along the path of the gas stream mark the distance from the white dwarf at intervals of $ 0.1R_{L1}$, ranging from 1.0$R_{L1}$ at the secondary star to 0.3$R_{L1}$ at the point of closest approach, where $R_{L1}$ is the distance from the white dwarf to the inner Lagrangian point. }
  \label{dopp2}
\end{figure}

In order to help determine the origin of the features in the trailed spectra, we constructed Doppler tomograms. Doppler tomography is an indirect imaging technique that maps emission/absorption features in the spectrum into velocity space. For a detailed review of Doppler tomography see 
\citet{marsh00}.
This technique assumes, amongst other things, that all points are equally visible at all times and that the flux from any point in the rotating frame is constant. These assumptions are violated to some extent in CI Aql, causing features to smear in the maps. Nevertheless, the location and relative strengths of the strongest features are reliably reproduced, as can be confirmed by inspecting the trailed spectra.
Fig. \ref{dopp2} shows Doppler maps for the He {\sevensize II} $\lambda4686$\AA\  and H$\alpha$ lines. 
The He {\sevensize II} emission appears to emanate from two principal areas, the disc and the centre of mass of the system, which is coincident with the irradiated face of the secondary. This explains the narrow constant velocity feature seen in the trailed spectra in Fig. \ref{tr}. The H$\alpha$ map contains a region of strong emission, indicated by darker grey-scales, which is located on the disc, and a region of strong absorption, indicated by lighter grey-scales, coincident with the secondary.

\subsection{Radial velocity of the white dwarf}
\label{rvwd}

\begin{figure}
  \vspace*{8pt}
\includegraphics[width=65mm,angle=270]{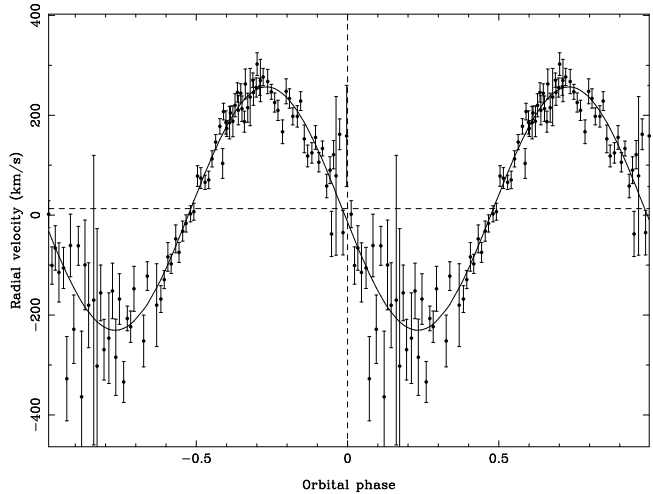}
  \caption{Radial-velocity curve of He {\sevensize II} $\lambda4686$\AA\ measured using a double-Gaussian fit with a separation of $ 900 $ km\,s$^{-1}$. The data have been folded for clarity. The horizontal dashed line represents the systemic velocity.}
  \label{rv}
\end{figure}

It is not possible to measure the radial velocity of the white dwarf directly. Instead, we rely on measuring the motion of the emission-line wings, since these are presumably formed in the inner parts of the accretion disc and should therefore reflect the motion of the white dwarf with the highest reliability. We used the double-Gaussian method of 
\citet{schneider80},
 since this technique is sensitive mainly to the motion of the line wings. The continuum-subtracted spectra were binned on to a constant velocity-interval scale about each of the emission-line rest wavelengths. The Gaussian widths were varied from 150 to 300~km\,s$^{-1}$
(FWHM) and we varied their separation from 400 to 1200~km\,s$^{-1}$. We then fitted the radial velocities, $v$, to the function

\begin{equation}
   v=\gamma-K \sin [2{\upi}(\phi-\phi_0)],
\end{equation}
where $\gamma$ is the systemic velocity, $K$ is the trial radial velocity semi-amplitude of the  white dwarf, $\phi$ is the orbital phase and $\phi_0$ is the offset between phase zero and the phase at which the radial velocity crosses from red to blue shifted. An example of a radial velocity curve for He {\sevensize II} $\lambda4686$\AA\ for a Gaussian width of 200~km\,s$^{-1}$ and separation of 900~km\,s$^{-1}$ is shown in Fig. \ref{rv}.

\begin{figure}
  \vspace*{8pt}
\includegraphics[width=65mm,angle=270]{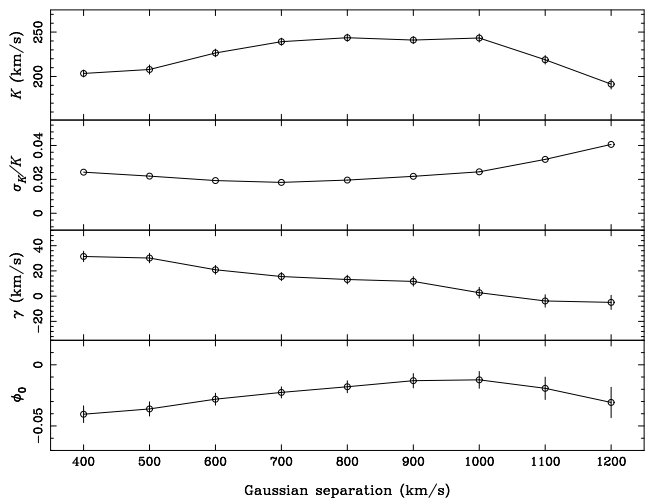}
  \caption{Diagnostic diagram for CI Aql based on the double-Gaussian radial-velocity fits to He {\sevensize II} $\lambda4686$\AA .}
  \label{diag}
\end{figure}

\begin{figure}
  \vspace*{8pt}
\includegraphics[width=65mm,angle=270]{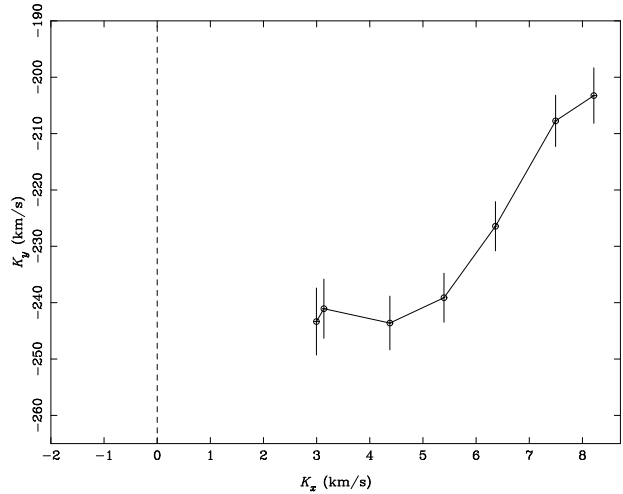}
  \caption{Light centre diagram for He {\sevensize II} $\lambda4686$\AA\ in CI Aql. The Gaussian separation of the points increases from right to left.}
  \label{lc1}
\end{figure}

The radial-velocity curves of the Balmer lines are contaminated by absorption from the secondary star and cannot be used as a reliable indicator of the motion of the white dwarf. The radial velocity curve of He {\sevensize II} $\lambda4686$\AA, on the other hand, has only a  negligible phase offset which suggests that it is a good representation of the radial velocity of the white dwarf, $K_\mathrm{W}$. In order to determine $K_\mathrm{W}$ we plotted a diagnostic diagram
\citep{shafter86}
 -- see Fig. \ref{diag}. $K$ approaches $K_\mathrm{W}$ as $\phi_0$ approaches zero, and the point at which the fractional error $\sigma_K/K$ starts to increase is taken as the optimum Gaussian separation. Fig. \ref{diag} shows that this occurs around a Gaussian separation of 900 km\,s$^{-1}$, giving a value of $K_\mathrm{W}=244 \pm 10$~km\,s$^{-1}$ and $\gamma=13 \pm 4$ km\,s$^{-1}$. Since the optimum value of $K_\mathrm{W}$ derived from the diagnostic diagram is signal-to-noise dependent, we also constructed a light-centre diagram
 \citep{marsh88a}
 -- see Fig. \ref{lc1}.  In the corotating coordinate frame, the white dwarf has a radial velocity of $(0,-K_\mathrm{W})$, and it can be seen that the largest Gaussian separations are almost coincident with this. Extrapolating the last point to the $K_y$ axis gives a radial velocity  for the white dwarf of $K_\mathrm{W}=244 \pm 10$ km\,s$^{-1}$, in agreement with the result from the diagnostic diagram.
  
\subsection{Radial velocity of secondary star}
\label{rad}
The calculation of the radial velocity, $K_\mathrm{R}$, the rotational velocity, $v \sin i$, and the spectral type and luminosity class of the secondary star requires an iterative process. An initial value of $K_\mathrm{R}$ is derived by cross-correlation and skew-mapping. To do this, we estimated a pair of initial values of $K_\mathrm{R}$ and $v \sin i$ to broaden the spectra to account for the radial and rotational velocities of the secondary. The resulting value of $K_\mathrm{R}$ is applied to the optimal subtraction technique, as described in Section \ref{rot}, which gives a revised value of $v \sin i$. This is then substituted back into the cross-correlation and the process is repeated until the values of $K_\mathrm{R}$ and $v \sin i$ converge.

\begin{figure}
  \vspace*{8pt}
\includegraphics[width=60mm,angle=270]{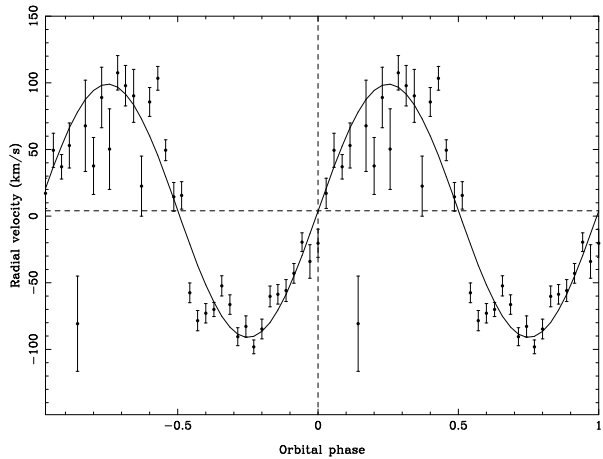}
  \caption{Radial-velocity curve of the secondary star in CI Aql, computed from the cross-correlation functions of the red spectra with an F8IV template.}
  \label{ccf}
\end{figure}

During our observations we had taken spectra of 53 template stars ranging from spectral types F8 to M7, and luminosity classes III to V. We performed iterations using the majority of the templates (we omitted some of the later-type templates when it became clear that the spectral type is quite early - see Section \ref{rot}). The secondary star in CI Aql is best observed through the absorption
lines seen in the trailed spectra of Fig. \ref{tr1}. 
We compared regions of the spectrum rich in absorption lines with the template stars, from which we had removed their radial velocities. We normalized each template spectrum
by dividing by its mean, and then subtracting a
higher-order spline fit to the continuum. This ensures that line strength is
preserved along the spectrum. The CI Aql spectra
were normalized in the same way. The smearing of the CI Aql spectra due to orbital motion during the exposure was always less than 5 km\,s$^{-1}$, which is insignificant compared to the rotational velocity, so no correction was required.  The template spectra were broadened by the rotational velocity of the secondary, $v \sin i$ (see Section \ref{rot}). Regions of the spectrum devoid of emission and contaminating absorption features e.g. Interstellar Medium (ISM),
 were then cross-correlated with each of the templates, yielding
a time series of cross-correlation functions (CCFs) for each template
star. We measured the position of the peak of each CCF using a parabolic approximation to three points around the maximum, and then fitted a sine curve to the resulting peaks. The lowest reduced $\chi^2$ (3.78) was found with an F8IV template in the red and the resulting radial velocity curve is shown in Fig. \ref{ccf}. This gives $\gamma=  42 \pm 3$ km\,s$^{-1}$ and $K_\mathrm{R} = 111 \pm 4$ km\,s$^{-1}$ . Other templates give values lying in the range $K_\mathrm{R} = 106-115$ km\,s$^{-1}$ in the red, and $K_\mathrm{R} = 96-100$ km\,s$^{-1}$ in the blue, with no clear trend with spectral type. The systemic velocity, $\gamma$, measured from the secondary star is slightly higher than that derived indirectly from the white dwarf via the He {\sevensize II} diagnostic diagram (see Section \ref{rvwd}); this may be due to systematic errors or evidence for a weak outflow in the He {\sevensize II} emitting region at the disc centre.

Due to noise in the individual CCFs, the radial velocity curves, such as the one shown in Fig. \ref{ccf}, are rather noisy and the derived value of $K_\mathrm{R}$ may be improved using the technique of skew mapping (see
\citealt{smith98} for details).

\begin{figure}
  \vspace*{8pt}
\includegraphics[width=130mm,angle=270]{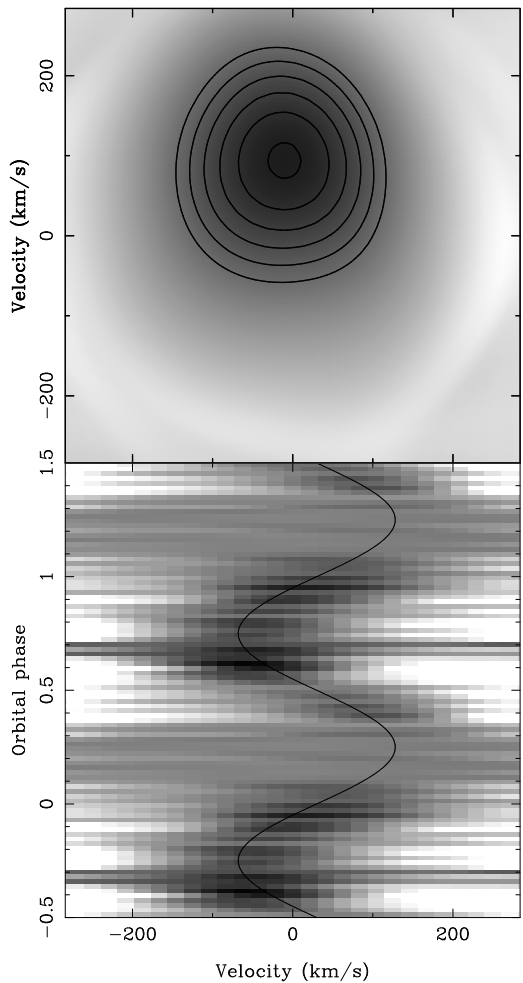}
  \caption{Skew map (top) and trailed CCFs (bottom) of CI Aql, computed in the red with an F8IV template. }
  \label{skew}
\end{figure}

To produce skew maps, the CCFs were back-projected in
the same way as time-resolved spectra in standard Doppler tomography
\citep{marsh88b}.
 If there is a detectable secondary star,
we expect a peak at (0, $K_\mathrm{R}$) in the skew map. This can be repeated
for each of the templates. Fig. \ref{skew} shows a red skew map produced by cross-correlating with an F8IV template star, together with the trailed CCFs. The peak of this skew map, which gave the strongest peak, gives a value of $K_\mathrm{R} = 105 \pm 5$ km\,s$^{-1}$, consistent with the radial-velocity curve technique described above. Other templates give peak values lying in the range $K_\mathrm{R} = 103-109$ km\,s$^{-1}$, with no clear trend with spectral type. Taking into account all of the uncertainties in the two techniques, we therefore adopt a value of  $K_\mathrm{R} = 105 \pm 10$ km\,s$^{-1}$.

The Doppler maps in Fig. \ref{dopp2} suggest that there is irradiation of the inner face of the secondary star by the accretion regions. This could affect the distribution of the absorption-line flux which would in turn skew the measured value of $K_\mathrm{R}$ from its true value (the radial velocity of the centre of mass of the secondary star). It is possible to correct $K_\mathrm{R}$ by modelling the distribution of absorption-line flux and comparing the resulting absorption-line light curves and radial-velocity curves with the observed data (see 
\citealt{thoroughgood04} 
for an example). Unfortunately the radial-velocity curve presented in Fig. \ref{ccf} is too noisy to distinguish a non-sinusoid caused by irradiation. The noise in the radial velocity curve dominates the errors in the mass determination compared to the likely shift due to irradiation. To double check this, we computed the absorption line flux as a function of orbital phase using optimal subtraction (see Section \ref{rot}).
The resulting light curve was extremely noisy and showed no evidence of variable absorption at phase 0.5, further supporting our conclusion that correcting our observations for irradiation is not appropriate.

\subsection{Rotational velocity and spectral type of secondary star}
\label{rot}
Using the value of $K_\mathrm{R}$ derived in Section \ref{rad}, we corrected the CI Aql spectra for orbital
motion and then averaged them. We rotationally broadened the spectral-type templates by a
range of velocities ($50-200$ km\,s$^{-1}$), using a linear limb-darkening coefficient of 0.5, and then performed optimal
subtraction
(\citealt{marsh94}).
 This technique subtracts a constant times the normalized
template spectrum from the normalized, average CI Aql
 spectrum, adjusting the constant to minimize the scatter in the residuals.
The scatter is computed from the $\chi^2$ between
the residual spectrum and a smoothed version of itself, avoiding regions of emission or contaminating absorption. By finding
the value of  rotational broadening that minimizes 
$\chi^2$ we obtain an estimate of both $v \sin i$ and the spectral type of
the secondary star. There should, strictly speaking, be a small
correction due to the intrinsic rotational velocity of the template
star, but as the templates are late-type stars they are assumed to be
sufficiently old to have lost most of their angular momentum by
magnetic braking and to have very small values of $v \sin i$ (of the order of
1 km\,s$^{-1}$;
\citealt{gray92}).

\begin{figure}
  \vspace*{8pt}
\includegraphics[width=70mm,angle=270]{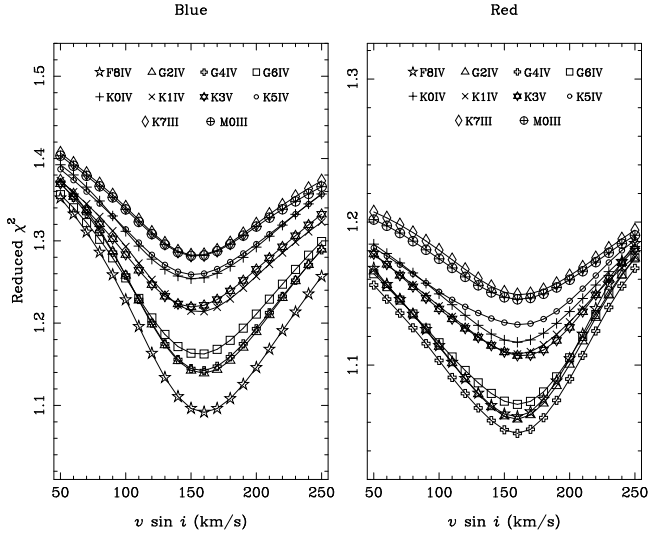}
  \caption{Reduced $\chi^2$ from the optimal subtraction technique plotted against $v \sin i$ for a range of templates in the blue and red wavelength ranges.}
  \label{optsub}
\end{figure}

Fig. \ref{optsub} shows the results of the optimal subtraction technique in the blue and red for a representative sample of the template stars. The minimum of the lowest curve gives both the value of $v \sin i$ and the spectral type of the secondary. Encouragingly, the two wavelength ranges and the different templates all give very similar minima, ranging from 
$v \sin i = 151-164$ km\,s$^{-1}$, with no clear trend with spectral type. In the blue the best fitting template is an F8IV star with 
$v \sin i = 160$ km\,s$^{-1}$, whilst in the red it is a G4IV star with $v \sin i = 159$ km\,s$^{-1}$, although the F8IV is almost equally preferred in the red with a value of $v \sin i = 157$ km\,s$^{-1}$. We therefore adopt the F8IV as the best-fitting template for our subsequent analysis and select a final value of $v \sin i = 158 \pm 7$ km\,s$^{-1}$.

The mass and radius of the secondary star are derived in Section \ref{par} using a Monte Carlo analysis. The results are consistent to within
the error bars with those of a slightly-evolved A0 star according to the
values published by 
\citet{gray92}.
 The presence of the Mg {\sevensize I}b, Ca {\sevensize I} and Fe {\sevensize I} absorption features also helps to constrain the spectral type of the secondary star, as these features disappear in stars earlier than A0 (see
 \citealt{pickles98}).
 Our optimal subtraction analysis shows that the secondary star is no later than F8. It is regrettable that we did not take spectra of templates earlier  than F8, which would have allowed us to constrain the spectral type more tightly.

\begin{figure}
  \vspace*{8pt}
\includegraphics[width=68mm,angle=270]{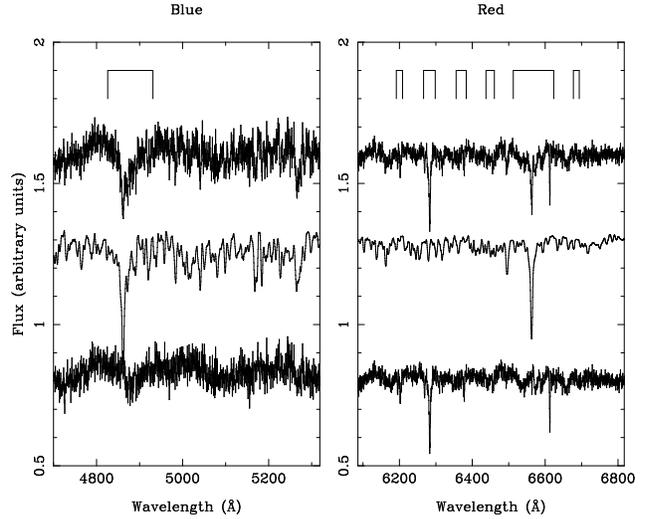}
  \caption{Orbitally--corrected spectra of CI Aql (top), the best-fitting F8IV template (middle) and the residuals after optimal subtraction (bottom). The brackets above the plots indicate the regions excluded from the optimal subtraction because they contain strong Balmer and ISM absorption features.}
  \label{temp}
\end{figure}

\begin{figure}
  \vspace*{8pt}
\includegraphics[width=65mm,angle=270]{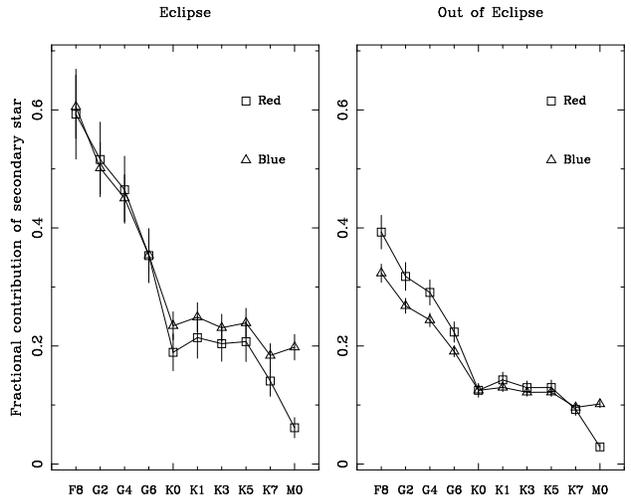}
  \caption{The fractional contribution of the secondary star to the total light in eclipse (left-hand panel) and out of eclipse (right-hand panel) in the red (squares) and blue (triangles). }
  \label{frac}
\end{figure}

\subsection{Distance to CI Aql}
\label{dis}

As noted above, we did not take spectra of templates earlier  than F8, which constrains our calculation of the distance since we can only place a lower limit using our best-fitting template (F8IV).

Fig. \ref{temp} shows the best-fitting template (F8IV), subtracted from the orbitally-corrected average spectrum of CI Aql. Some structure remains in the residuals probably due to the lack of an early enough template spectrum. The optimal subtraction technique also returns the value of the
constant by which the template spectra must be multiplied which, for
normalized spectra, is the fractional contribution of the secondary
star to the total light. The results are plotted in Fig. \ref{frac}, measured from the average CI Aql eclipse and non-eclipse spectra. 

By finding the apparent magnitude of the secondary star from its
contribution to the total light during eclipse, and estimating its
absolute magnitude, we can calculate the distance to CI Aql. The apparent magnitude can be found from the light curves (see Fig. \ref{lc}), by assuming that the red wavelength range is equivalent to the Johnson R-band and the blue range is equivalent to the V-band. At mid-eclipse, the flux in the red wavelength range is $1.4 \,\pm\, 0.1$ mJy, giving an apparent magnitude of $m_R=15.8 \pm 0.1$, whilst in the blue, the flux is $0.55 \pm 0.05$ mJy, giving $m_V=17.1 \pm 0.1$. In the red wavelength range the secondary star contributes $59\%$ and in the blue it contributes
$61\%$. This gives us approximate apparent magnitudes of
$m_R =16.4\pm 0.1$ and $m_V  = 17.6\pm 0.1$ for the secondary star.
Using the absolute magnitude of an F8IV star of $M_V=2.4$ (see
\citealt{gray09}),
the distance can be calculated with the distance modulus equation:

\begin{equation}
\label{dist}
5 \; $log$ (d/10)  = m_V - M_V - A_V,
\end{equation}
where $d$ is the distance to the star, and $A_V$ is the extinction in the $V$-band. The extinction to CI Aql was calculated by
\citet{lynch04}
to be $A_V =4.6 \pm 0.5$ using the intrinsic flux ratios of the Ly$\beta$-fluoresced O${\rm I}$ lines. We find a distance with this method of $d=1.3 \pm0.2$ kpc.

The distance can also be estimated using the Barnes-Evans relation 
(\citealt{barnes76}):

\begin{equation}
\label{bar}
F_{\nu} = 4.2207 - 0.1 V_0 - 0.5 $ log $ \phi= 3.977 - 0.429(V-R)_0,
\end{equation}
where $V_0$ and ($V-R$)$_0$ are the unreddened $V$ magnitude and $V-R$ colour of the secondary star, and $\phi=2R_2/d$ is the stellar angular diameter in milliarcseconds. We used a value of ($V-R$)$_0 = 0.37$ for an F8V star from
\citet{ducati01}, 
noting that luminosity class makes negligible difference. Using our measured value of $R_2= 2.07 \,\pm\, 0.06\, R_{\sun}$ (see Section \ref{par}), we find a distance of $1.2\pm0.1$ kpc.

Our distances are lower than other published estimates for CI Aql.  
\citet{lynch04}
 measured the distance to be $2.6  \pm  0.3$ kpc, using the relationship between the time taken to drop two magnitudes, $t_2$, and the absolute visual peak magnitude, $M_V$, of the 2000 outburst. 
 \citet{hachisu03}
 used theoretical models to derive a distance of 1.55 kpc. The distance was calculated by 
\citet{iijima12}
 as $2.1-2.7$ kpc, depending on the assumed value of the maximum luminosity of the 2000 outburst and using the extinction $A_V=4.6 \pm 0.5$ presented by 
\citet{lynch04}.
It should be stressed however, that our distance estimates are almost certainly lower limits as the spectral type of the secondary star is likely to be earlier than F8, in which case $M_V$ and $(V-R)_0$ decrease, thereby increasing the distance measurement. 

\subsection{System parameters}
\label{par}

Our measurements of the radial velocities of the white dwarf and secondary star, $K_\mathrm{W} = 244 \pm 10$ km\,s$^{_-1}$, $K_\mathrm{R} =105 \pm  10$ km\,s$^{-1}$, and the rotational broadening,
$v \sin i = 158 \pm 7$ km\,s$^{-1}$, can
now be used in conjunction with the period, $P$, and
our measurement of the eclipse full-width at half-depth,
$\Delta\phi_{1/2}= 0.09 \pm 0.01$ (an average value measured from the blue and red continuum light curves shown in Fig. \ref{lc}), to determine accurate system parameters
for CI Aql. Only four of these five measurements are
needed to calculate the system parameters. However, ignoring the error bars, we obtain different results for the system parameters depending on which set of four measurements are used. Hence, we have opted for a Monte Carlo
approach similar to that of 
\citet{horne93}.
For a given set of values of $K_\mathrm{R}$,
$v \sin i$, $\Delta \phi_{1/2}$  and $P$, the other system parameters are calculated
as follows.
$R_2/a$ can be estimated because the secondary star fills its Roche
lobe; $R_2$ is the equatorial radius of the secondary star and $a$ is the
binary separation. $R_2$ is approximately given by the volume-equivalent radius of the secondary star's Roche lobe, $R_V$, which is given by
\begin{equation}
\label{egg}
\frac{R_2}{a} \approx \frac{R_V}{a}= \frac{0.49q^{2/3}}{0.6q^{2/3}+{\rm ln}(1+q^{1/2})},
\end{equation}
where $q=M_2/M_1$ 
\citep{eggleton83}.
This expression is accurate to better than 1\% and is close to the equatorial radius of the secondary star as seen during eclipse.
The secondary star rotates synchronously with the orbital motion,
so we can combine $K_\mathrm{R}$ and $v\sin i$ to get

\begin{equation}
\label{synch}
\frac{R_2}{a}(1+q)=\frac{v \sin i}{K_\mathrm{R}}.
\end{equation}
This gives us two simultaneous equations so we can calculate the
mass ratio $q$ and $R_2/a$.
The orbital inclination, $i$, is uniquely determined by $q$ and $\Delta \phi_{1/2}$. We 
determine the inclination using a Roche lobe simulation similar to that of 
\citet{horne82}, 
which differs by $\sim 1$\% from the value derived using a simple geometric relation for the point eclipse by  a 
spherical body (see, for example,
 \citealt{dhillon92}).
  Newton's derivation of Kepler's Third Law gives us
\begin{equation}
\label{kep}
\frac{K_\mathrm{R}^3P}{2 \upi G} = \frac{M_1\sin^3i}{(1+q)^2},
\end{equation}
which, with the values of $q$ and $i$ calculated above, 
gives the mass of the primary star. The mass of the
secondary star can then be obtained using the mass ratio $q$. 
The radius of the secondary star, $R_2$, is obtained from the equation
\begin{equation}
\label{vs}
v \sin i = \frac{2 \upi R_2 \sin i}{P},
\end{equation}
and $a$ is calculated from
Equation (\ref{synch}) with $q$ and $i$ now known.

The Monte Carlo simulation takes 30,000 sample values of $K_\mathrm{R}$,
$v \sin i$ and $\Delta \phi_{1/2}$, treating each as being normally distributed about
their measured values with standard deviations equal to the errors
on the measurements. We treat the period, $P$, as a constant since it has been measured to a far higher degree of accuracy. We then calculate $M_1, M_2, i, R_2,$ and $a$, as outlined above, rejecting any 
($K_\mathrm{R}$, $v \sin i$, $\Delta \phi_{1/2}$) triplets which give $\sin i$ greater than $1$.
Each solution is also subject to rejection if the calculated value of $K_\mathrm{W}$ is inconsistent with our measured value of
$K_\mathrm{W}= 244 \pm 10$ km\,s$^{-1}$ derived in Section \ref{rvwd}. The probability of rejection is determined in accordance with a
Gaussian probability law, i.e. if the calculated value of $K_\mathrm{W}$ exactly equals our measured value then it is 100\% certain to be accepted, if it lies within 1$\sigma$
 of our measured value of $K_\mathrm{W}$ (with $\sigma$ equal to our measured error on $K_\mathrm{W}$), the probability of rejection is 68 per cent, at 2$\sigma$ it is 95 per cent, etc.
 
\begin{figure*}
  \vspace*{8pt}
\includegraphics[width=100mm,angle=270]{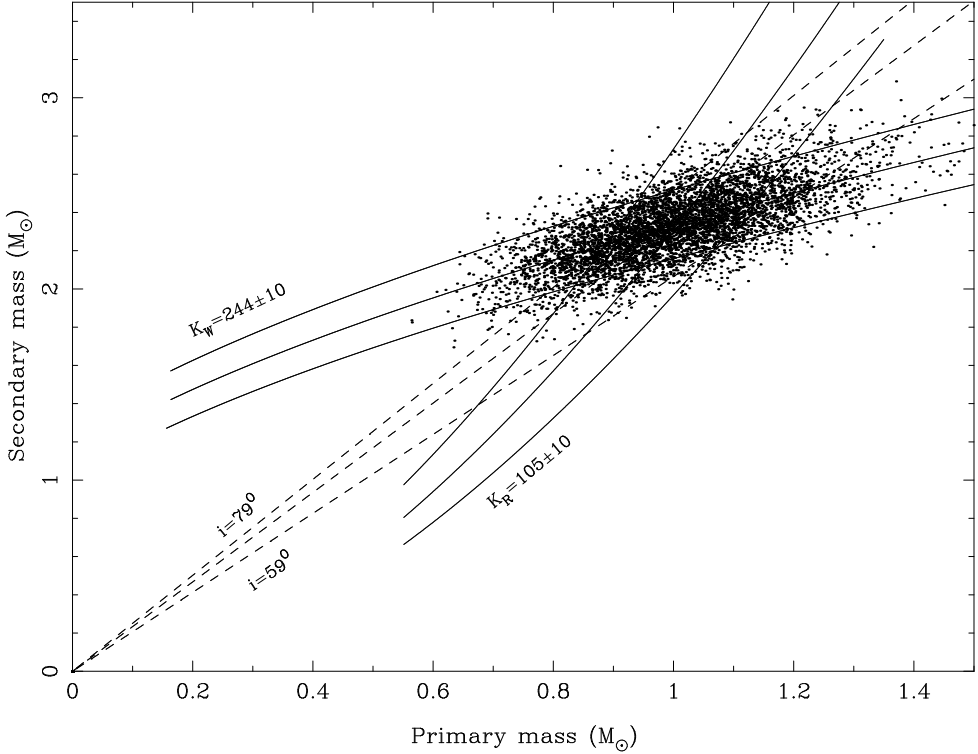}
  \caption{Results of the Monte Carlo system parameter determination for CI Aql. Each dot represents an $(M_1,M_2)$ pair. The dashed lines show constant inclinations of $i=59^{\circ},69^{\circ},79^{\circ}$. The solid curves satisfy the observed constraints  on $K_\mathrm{W}= 244 \pm 10$ km\,s$^{-1}$ and $K_\mathrm{R}=105 \pm 10$ km\,s$^{-1}$.}
  \label{mc}
\end{figure*}

 Each accepted $M_1$,$M_2$
pair is then plotted as a point in Fig. \ref{mc}, and the masses and their
errors are computed from the mean and standard deviation of the
distribution of  points. The solid curves in Fig. \ref{mc} satisfy the white dwarf radial-velocity constraint, $K_\mathrm{W}= 244 \pm 10$ km\,s$^{-1}$ and the
secondary star radial velocity constraint, $K_\mathrm{R}=105 \pm 10$ km\,s$^{-1}$.
We find that $M_1=1.00 \pm 0.14$ M$_{\sun}$ and $M_2=2.32  \pm 0.19$ M$_{\sun}$. The full results of the Monte Carlo simulations are shown in Table \ref{tab1}. Note that computing the system parameters without the $K_\mathrm{W}$ constraint gives very similar results and a derived value of $K_\mathrm{W}=246 \pm 10$ km\,s$^{-1}$, giving us confidence that the assumptions and techniques detailed in Section \ref{rvwd} are valid.

\begin{table}
  \begin{center}
  \caption{System parameters for CI Aql and their associated statistical errors.}
  \label{tab1}
  \begin{tabular}{|l|c|c|}\hline
{Parameter} & {Measured} & {Monte Carlo}  \\
 & {value} & {value} \\
\hline
 $\gamma$ (km\,s$^{-1}$) & $42 \pm 3$  \\
$K_\mathrm{W}$ (km\,s$^{-1}$) & $244 \pm 10$ & $244 \pm 7$  \,\, \\
$K_\mathrm{R}$ (km\,s$^{-1}$) & \,\,$105 \pm 10 $ $ $  \\
$v$ sin $i$ (km\,s$^{-1}$) &$158 \pm 7$ $ $  \\
$\Delta\phi_{1/2}$ & \,\, $0.09 \pm 0.01$ \\
$P$ (days) & $0.6183634(3)$  \\
$i$\,($^{\circ}$) & & $69 \pm 2$ \, \\
$M_{1} \,(M_{\sun})$ & & $1.00 \pm 0.14$ \\
$M_2 \, (M_{\sun})$ & & $2.32 \pm 0.19$ \\
$q$ & & $2.35 \pm 0.24$ \\
$R_2$ $(R_{\sun})$ & & $2.07 \pm 0.06$  \\
$a$ $(R_{\sun})$ & & $4.56 \pm 0.14$  \\
$R_\mathrm{D}$ $(R_{L1})$ & & $0.45 \pm 0.10$  \\
Distance (kpc) & $1.3 \pm 0.2$\\
\hline
\end{tabular}
\end{center}
\end{table}

We also estimated the radius of the accretion disc using the half-width of the top of the eclipse (i.e. timing the first and last contacts of eclipse and dividing by 2),  $\Delta \phi$, measured from our light curve and those given by
\citet{schaefer11}. The radius of the disc is given by
\begin{equation}
\label{drad}
R_\mathrm{D}= a \sin i (2 \upi \Delta \phi) - R_\mathrm{C},
\end{equation}
where $R_\mathrm{C}$ is the half-chord on the secondary, given by
\begin{equation}
\label{rc}
R_\mathrm{C}=a \,\,\sqrt{\left[\left(\frac{R_2}{a}\right)^2 - \cos^2 i\right]}.
\end{equation}
Using our measured value of $\Delta \phi = 0.08 \pm 0.02$, and $q$ and $i$ derived above, we calculate the disc radius $R_\mathrm{D}$= 0.45 $\pm$ 0.10 $R_{L1}$, where $R_{L1}$ is the distance from the WD to the inner Lagrangian point.

One of the assumptions that we have made in the above calculations is that the measured $\Delta \phi_{1/2}$ corresponds to the phases when the Roche lobe of the secondary star eclipses the centre of the disc, i.e. that $-\frac{\Delta \phi_{1/2}}{2}$ and
$+\frac{\Delta \phi_{1/2}}{2}$ represent the start and end of the eclipse of the centre of the disc. We can verify that the centre of
the accretion disc is indeed eclipsed in CI Aql as follows. Assuming an axisymmetric brightness distribution on the accretion disc, which seems reasonable given the lack of evidence for a bright spot, if the flux from the white dwarf/disc component drops by more than half at mid eclipse then the centre of the disc must be eclipsed. From Fig. \ref{lc} we can see that the blue out-of-eclipse average flux is approximately 0.92\,mJy, whereas the minimum flux during eclipse is approximately 0.56\,mJy. Our measured value of the contribution of the secondary out of eclipse from Fig. \ref{frac} is 32\% and in eclipse it is 61\%. Therefore the flux from the white dwarf/disc component out of eclipse is 0.63\,mJy whereas in eclipse the flux reduces to 0.22\,mJy. Hence the flux from the white dwarf/disc component reduces by 65\% during eclipse, thereby demonstrating that the centre of the disc is indeed eclipsed. We also need to verify that the eclipse of the disc centre commences at phase $-\frac{\Delta \phi_{1/2}}{2}$ and ends at phase $+\frac{\Delta \phi_{1/2}}{2}$, i.e. that the brightness profile of the disc, the curved limb of the secondary star and the uneclipsed portions of the disc do not skew our measurement of $\Delta \phi_{1/2}$. We therefore ran a series of light-curve simulations using the CI Aql system parameters, where a Roche lobe eclipses a disc with a range of different radii and axisymmetric brightness profiles. We found negligible variation in the value of $\Delta \phi_{1/2}$ measured from these simulated light curves. For clarity, we should stress that the error on $i$ quoted in Table \ref{tab1} is purely statistical, derived from the mean and standard deviation of the output from the Monte Carlo simulation, and hence does not include the effect of any systematic errors introduced by, for example, the assumption of an axisymmetric disc brightness distribution.

Our measurement of the inclination compares well with the value of $i=71^{\circ}$ found by \citet{lederle03}, who used a 3D model of CI Aql to fit its light curve. Their slightly higher inclination is most likely due to the use of a smaller assumed mass, and hence radius, of the secondary in CI Aql.

\section{Discussion}
 
There are two prerequisites for an RN, a high-mass white dwarf and a high mass-transfer rate ($\dot{M}$; 
\citealt{anupama08}).
 We have found that CI Aql  does indeed have a relatively high mass  white dwarf ($M_{1} =1.00 \pm 0.14$ M$_{\sun}$).
 \citet{hachisu03}
calculated $\dot{M}$ of CI Aql by estimating the mass ejected during the 2000 outburst. Assuming that there were no other outbursts since 1917, their model showed the rate to be $\dot{M}=1\times10^{-7} $M$_{\sun}$ yr$^{-1}$. If there was another outburst in 1947, as suggested by 
\citet{schaefer01},
then the rate increases  to $\dot{M}=1.3\times10^{-7} $M$_{\sun}$ yr$^{-1}$. This is much higher than the average mass-transfer rates in CVs, which are of  order $10^{-8}-10^{-9} $M$_{\sun}$ yr$^{-1}$ (see
\citealt{knigge11a}
for a review).

 Mass transfer can be driven by either angular momentum loss or the evolution of the secondary. The models of
\citet{langer00},
\citet{han04}
and
\citet{hachisu08a}
suggest that, in systems with the parameters of CI Aql, it is the evolution of the secondary that is the key contributor to the high $\dot{M}$. In these models, the evolutionary expansion brings the star into contact with its Roche lobe and, because $M_2 > M_1$,  thermal-timescale mass transfer begins and the system exhibits nova outbursts.
Eventually, the nuclear evolution of the star causes it to expand further, thereby increasing the mass-transfer rate, which leads to steady hydrogen burning on the surface of the white dwarf. Such systems are supersoft X-ray sources
\citep{shen07}.
The white dwarf mass grows considerably during this phase, ultimately exploding as an SNIa.

The zero-age main-sequence radius of a 2.3 M$_{\sun}$ star is 1.8 R$_{\sun}$ and increases to our observed value of 2.07 R$_{\sun}$ after approximately 300 Myr, according to the models of
\citet{langer00}.
This suggests that the secondary star in CI Aql is a core hydrogen burning star that has begun to evolve away from the main sequence towards the giant branch. This expansion has brought it into contact with its Roche lobe, and thermal-timescale mass transfer has recently started.

We note that our measured values for CI Aql, when plotted on fig. 1 of
\citet{hachisu08},
places it in the region of a MS + WD binary system that  will evolve to produce a SNIa. An approximate estimate of the time until the system explodes as a SNIa can be found using the models of 
\citet{langer00},
which suggest that CI Aql will explode as a SNIa in less than 10 Myr. Similarly, 
\citet{ivanova04}
developed numerical models of the evolution of binaries consisting of evolved MS stars and high mass WD companions, that are undergoing thermal-timescale mass transfer. Their models show the parameters of WD+MS binaries that produce SNIa and are consistent with those of 
\citet{hachisu08a}. 
Fig. 1 from
\citet{ivanova04}
shows the evolutionary fates of binary systems that begin mass transfer with various initial WD and secondary star masses. The parameters of CI Aql place it in the region that produces SNIa.

\section{Conclusions}

We have shown that CI Aql contains a high-mass white dwarf of $M_1$=1.00 $\pm$ 0.14 M$_{\sun}$.  The secondary star has a mass of $M_2$=2.32  $\pm$ 0.19 M$_{\sun}$ and a radius of $R_2$=2.07 $\pm$ 0.06 R$_{\sun}$, implying it is a slightly evolved early A-type star.

We have compared our system parameters with theoretical models of the evolution of binary systems. CI Aql is currently in the RN phase, and is expected to evolve into a supersoft X-ray source. Models suggest it may ultimately explode as a SNIa, possibly within 10 Myr.

\section*{Acknowledgement}

The WHT is operated on the island of La Palma by the Isaac Newton Group in the Spanish Observatorio del Roque de los Muchachos of the Instituto de Astrofisica de Canarias. We thank the anonymous referee for helpful comments on the manuscript. 

\bibliographystyle{mn2e.bst}
\bibliography{abbrev.bib,refs.bib}

\begin{thebibliography}{}

\bibitem[\protect\citeauthoryear{Anupama}{Anupama}{2008}]{anupama08}
Anupama G.~C.,  2008, PASP, 401, 31

\bibitem[\protect\citeauthoryear{Badenes, Hughes, Bravo \& Langer}{Badenes
  et~al.}{2007}]{badenes07}
Badenes C.,  Hughes J.~P.,  Bravo E.,    Langer N.,  2007, ApJ, 662, 472

\bibitem[\protect\citeauthoryear{Badenes \& Maoz}{Badenes \&
  Maoz}{2012}]{badenes12}
Badenes C.,  Maoz D.,  2012, ApJ, 749, L11

\bibitem[\protect\citeauthoryear{Barnes \& Evans}{Barnes \&
  Evans}{1976}]{barnes76}
Barnes T.,  Evans D.,  1976, MNRAS, 174, 489

\bibitem[\protect\citeauthoryear{{Bloom}, {Kasen}, {Shen}, {Nugent}, {Butler},
  {Graham}, {Howell}, {Kolb}, {Holmes}, {Haswell}, {Burwitz}, {Rodriguez} \&
  {Sullivan}}{{Bloom} et~al.}{2012}]{bloom12}
{Bloom} J.~S.,  et al.,  2012, ApJ, 744, L17

\bibitem[\protect\citeauthoryear{{Brown}, {Dawson}, {de Pasquale}, {Gronwall},
  {Holland}, {Immler}, {Kuin}, {Mazzali}, {Milne}, {Oates} \& {Siegel}}{{Brown}
  et~al.}{2012}]{brown12}
{Brown} P.~J.,  et al.,  2012, ApJ, 753, 22

\bibitem[\protect\citeauthoryear{Carter, Benn, Rutten, Breare, Rudd, King,
  Clegg, Dhillon, Arribas, Rasilla, Garcia, Jenkins \& Charles}{Carter
  et~al.}{1993}]{carter93}
Carter D.,  et al.,  1993, User Manual~24, WHT -- ISIS
  Users' Manual.
Isaac Newton Group, La Palma

\bibitem[\protect\citeauthoryear{Claret \& Bloemen}{Claret \&
  Bloemen}{2011}]{claret11}
Claret A.,  Bloemen S.,  2011, A\&A, 529, A75

\bibitem[\protect\citeauthoryear{Dhillon, Marsh, Jones \& Smith}{Dhillon
  et~al.}{1992}]{dhillon92}
Dhillon V.~S.,  Marsh T.~R.,  Jones D. H.~P.,    Smith R.~C.,  1992, MNRAS,
  258, 225

\bibitem[\protect\citeauthoryear{Ducati, Bevilacqua, Rembold \& Ribeiro}{Ducati
  et~al.}{2001}]{ducati01}
Ducati J.,  Bevilacqua C.,  Rembold S.,    Ribeiro D.,  2001, ApJ, 558, 309

\bibitem[\protect\citeauthoryear{Duerbeck}{Duerbeck}{1987}]{duerbeck87a}
Duerbeck H.~W.,  1987, Space Sci.~Rev., 45, 1

\bibitem[\protect\citeauthoryear{Eggleton}{Eggleton}{1983}]{eggleton83}
Eggleton P.~P.,  1983, ApJ, 268, 368

\bibitem[\protect\citeauthoryear{Gray}{Gray}{1992}]{gray92}
Gray D.~F.,  1992, The Observation and Analysis of Stellar Photospheres.
Cambridge University Press, Cambridge

\bibitem[\protect\citeauthoryear{Gray \& Corbally}{Gray \&
  Corbally}{2009}]{gray09}
Gray R.~O.,  Corbally C.~J.,  2009, Stellar Spectral Classification.
Princeton University Press, Princeton, NJ

\bibitem[\protect\citeauthoryear{Greiner, Alcala \& Wenzel}{Greiner
  et~al.}{1996}]{greiner96}
Greiner J.,  Alcala J.,    Wenzel W.,  1996, Inf.\ Bull.\ Var.\ Stars, 4338, 1

\bibitem[\protect\citeauthoryear{Hachisu, Kato \& Nomoto}{Hachisu
  et~al.}{1999}]{hachisu99a}
Hachisu I.,  Kato M.,    Nomoto K.,  1999, ApJ, 522, 487

\bibitem[\protect\citeauthoryear{Hachisu, Kato \& Schaefer}{Hachisu
  et~al.}{2003}]{hachisu03}
Hachisu I.,  Kato M.,    Schaefer B.,  2003, ApJ, 584, 1008

\bibitem[\protect\citeauthoryear{Hachisu, Kato \& Nomoto}{Hachisu
  et~al.}{2008a}]{hachisu08a}
Hachisu I.,  Kato M.,    Nomoto K.,  2008a, ApJ, 679, 1390

\bibitem[\protect\citeauthoryear{Hachisu, Kato \& Nomoto}{Hachisu
  et~al.}{2008b}]{hachisu08}
Hachisu I.,  Kato M.,    Nomoto K.,  2008b, ApJ, 683, 127

\bibitem[\protect\citeauthoryear{{Hamuy}, {Phillips}, {Suntzeff}, {Maza},
  {Gonz{\'a}lez}, {Roth}, {Krisciunas}, {Morrell}, {Green}, {Persson} \&
  {McCarthy}}{{Hamuy} et~al.}{2003}]{hamuy03}
{Hamuy} M.,  et al.,  2003, Nat, 424, 651

\bibitem[\protect\citeauthoryear{Han \& Podsiadlowki}{Han \&
  Podsiadlowki}{2004}]{han04}
Han Z.,  Podsiadlowki P.,  2004, MNRAS, 350, 1301

\bibitem[\protect\citeauthoryear{Horesh, Kulkarni, Fox, Carpenter, Kasliwal,
  Ofek, Quimby, {Gal-Yam}, Cenko, {de Bruyn}, Kamble, Wijers, {van der Horst},
  Kouveliotou, Podsiadlowski, Sullivan, Maguire, Howell, Nugent, Gehrels \&
  Others}{Horesh et~al.}{2012}]{horesh12}
Horesh A.,  et al.,  2012, ApJ,
  746, 21

\bibitem[\protect\citeauthoryear{Horne, Lanning \& Gomer}{Horne
  et~al.}{1982}]{horne82}
Horne K.,  Lanning H.~H.,    Gomer R.~H.,  1982, ApJ, 252, 681

\bibitem[\protect\citeauthoryear{Horne, Welsh \& Wade}{Horne
  et~al.}{1993}]{horne93}
Horne K.,  Welsh W.~F.,    Wade R.~A.,  1993, ApJ, 410, 357

\bibitem[\protect\citeauthoryear{Iben \& Tutukov}{Iben \&
  Tutukov}{1984}]{iben84}
Iben I.,  Tutukov A.~V.,  1984, ApJS, 54, 335

\bibitem[\protect\citeauthoryear{Iijima}{Iijima}{2012}]{iijima12}
Iijima T.,  2012, A\&A, 544, A26

\bibitem[\protect\citeauthoryear{Ivanova \& Taam}{Ivanova \&
  Taam}{2004}]{ivanova04}
Ivanova N.,  Taam R.~E.,  2004, ApJ, 601, 1058

\bibitem[\protect\citeauthoryear{{Kerzendorf}, {Schmidt}, {Asplund}, {Nomoto},
  {Podsiadlowski}, {Frebel}, {Fesen} \& {Yong}}{{Kerzendorf}
  et~al.}{2009}]{kerzendorf09}
{Kerzendorf} W.~E.,  {Schmidt} B.~P.,  {Asplund} M.,  {Nomoto} K.,
  {Podsiadlowski} P.,  {Frebel} A.,  {Fesen} R.~A.,    {Yong} D.,  2009, ApJ,
  701, 1665

\bibitem[\protect\citeauthoryear{Knigge, Baraffe \& Patterson}{Knigge
  et~al.}{2011}]{knigge11a}
Knigge C.,  Baraffe I.,    Patterson J.,  2011, ApJS, 194, 28

\bibitem[\protect\citeauthoryear{Langer, Deutschmann, Wellstein \&
  H{\"o}flich}{Langer et~al.}{2000}]{langer00}
Langer N.,  Deutschmann A.,  Wellstein S.,    H{\"o}flich P.,  2000, A\&A, 362,
  1046

\bibitem[\protect\citeauthoryear{Lederle \& Kimeswenger}{Lederle \&
  Kimeswenger}{2003}]{lederle03}
Lederle C.,  Kimeswenger S.,  2003, A\&A, 397, 951

\bibitem[\protect\citeauthoryear{Leonard}{Leonard}{2007}]{leonard07}
Leonard D.~C.,  2007, ApJ, 670, 1275

\bibitem[\protect\citeauthoryear{Livio \& Riess}{Livio \&
  Riess}{2003}]{livio03}
Livio M.,  Riess A.,  2003, ApJ, 594, L94

\bibitem[\protect\citeauthoryear{Lynch, Wilson, Rudy, Venturini, Mazuk, Miller
  \& Puetter}{Lynch et~al.}{2004}]{lynch04}
Lynch D.,  Wilson J.~C.,  Rudy R.~J.,  Venturini C.,  Mazuk S.,  Miller N.~A.,
    Puetter R.~C.,  2004, AJ, 127, 1089

\bibitem[\protect\citeauthoryear{Maoz \& Mannucci}{Maoz \&
  Mannucci}{2012}]{maoz11}
Maoz D.,  Mannucci F.,  2012, PASA, 29, 447

\bibitem[\protect\citeauthoryear{Marsh}{Marsh}{1988}]{marsh88a}
Marsh T.~R.,  1988, MNRAS, 231, 1117

\bibitem[\protect\citeauthoryear{Marsh}{Marsh}{2000}]{marsh00}
Marsh T.~R.,  2000, in Boffin H.,  Steeghs D.,   Cuypers J.,  eds,
  Astrotomography - Indirect Imaging Methods in Observational Astronomy
  Springer-Verlag, Berlin

\bibitem[\protect\citeauthoryear{Marsh \& Horne}{Marsh \&
  Horne}{1988}]{marsh88b}
Marsh T.~R.,  Horne K.,  1988, MNRAS, 235, 269

\bibitem[\protect\citeauthoryear{Marsh, Robinson \& Wood}{Marsh
  et~al.}{1994}]{marsh94}
Marsh T.~R.,  Robinson E.~L.,    Wood J.~H.,  1994, MNRAS, 266, 137

\bibitem[\protect\citeauthoryear{Morris \& Naftilan}{Morris \&
  Naftilan}{1993}]{morris93}
Morris S.~L.,  Naftilan S.~A.,  1993, AJ, 419, 344

\bibitem[\protect\citeauthoryear{{Napiwotzki}, {Karl}, {Nelemans}, {Yungelson},
  {Christlieb}, {Drechsel}, {Heber}, {Homeier}, {Koester}, {Leibundgut},
  {Marsh}, {Moehler}, {Renzini} \& {Reimers}}{{Napiwotzki}
  et~al.}{2007}]{napiwotzki07}
{Napiwotzki} R.,  et al.,  2007, in
  Napiwotzki R.,  Burleigh M.,  eds, ASP Conf. Ser. 372, 15th European Workshop
  on White Dwarfs, Astron. Soc. Pac., San Francisco, p.~387

\bibitem[\protect\citeauthoryear{{Nugent}, {Sullivan}, {Cenko}, {Thomas},
  {Kasen}, {Howell}, {Bersier}, {Bloom}, {Kulkarni}, {Kandrashoff},
  {Filippenko}, {Silverman}, {Marcy}, {Howard}, {Isaacson}, {Maguire}, {Suzuki}
  \& Others}{{Nugent} et~al.}{2011}]{nugent11}
{Nugent} P.~E.,  et al.,
   2011, Nat, 480, 344

\bibitem[\protect\citeauthoryear{{Pakmor}, {Kromer}, {Taubenberger} \&
  {Springel}}{{Pakmor} et~al.}{2013}]{pakmor13}
{Pakmor} R.,  {Kromer} M.,  {Taubenberger} S.,    {Springel} V.,  2013, ArXiv
  e-prints, 1302.2913

\bibitem[\protect\citeauthoryear{{Perlmutter}, {Aldering}, {Goldhaber}, {Knop},
  {Nugent}, {Castro}, {Deustua}, {Fabbro}, {Goobar}, {Groom}, {Hook}, {Kim},
  {Kim}, {Lee}, {Nunes}, {Pain}, {Pennypacker}, {Quimby}, {Lidman} \&
  Others}{{Perlmutter} et~al.}{1999}]{perlmutter99}
{Perlmutter} S., et al.,
  1999, ApJ, 517, 565

\bibitem[\protect\citeauthoryear{Phillips}{Phillips}{1993}]{phillips93}
Phillips M.~M.,  1993, ApJ, 413, L105

\bibitem[\protect\citeauthoryear{Pickles}{Pickles}{1998}]{pickles98}
Pickles A.~J.,  1998, PASP, 110, 863

\bibitem[\protect\citeauthoryear{{Reinmuth}}{{Reinmuth}}{1925}]{reinmuth}
{Reinmuth} K.,  1925, Astron.~Nachr., 225, 385

\bibitem[\protect\citeauthoryear{Riess, Filippenko, Challis, Clocchiatti,
  Diercks, Garnavich, Gilliland, Hogan, Jha, Kirshner, Leibundgut, Phillips,
  Reiss, Schmidt, Schommer, Smith, Spyromilio, Stubbs, Suntzeff \& Tonry}{Riess
  et~al.}{1998}]{riess98}
Riess A.,  et al.,  1998, AJ, 116, 1009

\bibitem[\protect\citeauthoryear{Schaefer}{Schaefer}{2001}]{schaefer01}
Schaefer B.~E.,  2001, IAU Circ., 7750, 2

\bibitem[\protect\citeauthoryear{Schaefer}{Schaefer}{2010}]{schaefer10}
Schaefer B.~E.,  2010, ApJS, 187, 275

\bibitem[\protect\citeauthoryear{Schaefer}{Schaefer}{2011}]{schaefer11}
Schaefer B.~E.,  2011, ApJ, 742, 112

\bibitem[\protect\citeauthoryear{Schaefer \& Pagnotta}{Schaefer \&
  Pagnotta}{2012}]{schaefer12}
Schaefer B.~E.,  Pagnotta A.,  2012, Nat, 481, 164

\bibitem[\protect\citeauthoryear{Schneider \& Young}{Schneider \&
  Young}{1980}]{schneider80}
Schneider D.~P.,  Young P.~J.,  1980, ApJ, 238, 946

\bibitem[\protect\citeauthoryear{Shafter, Szkody \& Thorstensen}{Shafter
  et~al.}{1986}]{shafter86}
Shafter A.~W.,  Szkody P.,    Thorstensen J.~R.,  1986, ApJ, 308, 765

\bibitem[\protect\citeauthoryear{{Shen} \& {Bildsten}}{{Shen} \&
  {Bildsten}}{2007}]{shen07}
{Shen} K.~J.,  {Bildsten} L.,  2007, ApJ, 660, 1444

\bibitem[\protect\citeauthoryear{{Shen}, {Bildsten}, {Kasen} \&
  {Quataert}}{{Shen} et~al.}{2012}]{shen12}
{Shen} K.~J.,  {Bildsten} L.,  {Kasen} D.,    {Quataert} E.,  2012, ApJ, 748,
  35

\bibitem[\protect\citeauthoryear{{Shen}, {Guillochon} \& {Foley}}{{Shen}
  et~al.}{2013}]{shen13}
{Shen} K.~J.,  {Guillochon} J.,    {Foley} R.~J.,  2013, ArXiv e-prints,
  1302.2916

\bibitem[\protect\citeauthoryear{{Sim}, {Fink}, {Kromer}, {R{\"o}pke}, {Ruiter}
  \& {Hillebrandt}}{{Sim} et~al.}{2012}]{sim12}
{Sim} S.~A.,  {Fink} M.,  {Kromer} M.,  {R{\"o}pke} F.~K.,  {Ruiter} A.~J.,
  {Hillebrandt} W.,  2012, MNRAS, 420, 3003

\bibitem[\protect\citeauthoryear{{Simon}, {Gal-Yam}, {Gnat}, {Quimby},
  {Ganeshalingam}, {Silverman}, {Blondin}, {Li}, {Filippenko}, {Wheeler},
  {Kirshner}, {Patat}, {Nugent}, {Foley}, {Vogt}, {Butler} \& Others}{{Simon}
  et~al.}{2009}]{simon09}
{Simon} J.~D., et al .,  2009, ApJ, 702, 1157

\bibitem[\protect\citeauthoryear{Smith, Dhillon \& Marsh}{Smith
  et~al.}{1998}]{smith98}
Smith D.~A.,  Dhillon V.~S.,    Marsh T.~R.,  1998, MNRAS, 296, 465

\bibitem[\protect\citeauthoryear{Takamizawa, Kato \& Yamamoto}{Takamizawa
  et~al.}{2000}]{takamizawa00}
Takamizawa K.,  Kato T.,    Yamamoto M.,  2000, IAU Circ., 7409, 1

\bibitem[\protect\citeauthoryear{Thoroughgood, {Dhillon} \&
  {Littlefair}}{Thoroughgood et~al.}{2001}]{thoroughgood01}
Thoroughgood T.~D.,  {Dhillon} V.,    {Littlefair} S.,  2001, MNRAS, 327, 1323

\bibitem[\protect\citeauthoryear{Thoroughgood, {Dhillon}, {Watson}, {Buckley},
  {Steeghs} \& {Stevenson}}{Thoroughgood et~al.}{2004}]{thoroughgood04}
Thoroughgood T.~D.,  {Dhillon} V.,  {Watson} C.,  {Buckley} D.,  {Steeghs} D.,
    {Stevenson} M.,  2004, MNRAS, 353, 1135

\bibitem[\protect\citeauthoryear{van Kerkwijk, Chang \& Justham}{van Kerkwijk
  et~al.}{2010}]{vankerkwijk10}
van Kerkwijk M.~H.,  Chang P.,    Justham S.,  2010, ApJ, 722, 157

\bibitem[\protect\citeauthoryear{Webbink}{Webbink}{1984}]{webbink84}
Webbink R.,  1984, ApJ, 277, 355

\bibitem[\protect\citeauthoryear{Whelan \& Iben}{Whelan \&
  Iben}{1973}]{whelan73}
Whelan J.,  Iben I.,  1973, ApJ, 186, 1007

\bibitem[\protect\citeauthoryear{Yamaoka \& Shirakami}{Yamaoka \&
  Shirakami}{2000}]{yamaoka00}
Yamaoka H.,  Shirakami K.,  2000, IAU Circ., 7411, 1

\end{thebibliography}

\bsp

\label{lastpage}

\end{document}